# Current status of the High-Resolution Multi-Object Spectrograph (MOS-HR) for the Wide-field Spectroscopic Telescope

Andrea Tozzi[a*], Anna Brucalassi[a], Matteo Munari[a], Simone D'Auria[a], Andrea Bianco[b], Giorgio Pariani[b], Ciro Del Vecchio[a], Paolo Picchi[a], Roland Bacon[c], Bernard Delabre[d], David Lee[e], Sofia Randich[a], Will Saunders[d].

*[a]INAF – Osservatorio Astrofisico di Arcetri, Largo E. Fermi 5, 50125 Firenze, Italy; [b]INAF-Merate, Via E. Bianchi, 46, 23807 Merate LC; [c]CRAL, Université Claude Bernard Lyon 1, 69230 Saint-Genis-Laval, France; [d]Independent consultant; [e]UK Astronomy Technology Centre (UKATC/STFC), Royal Observatory Edinburgh, EH9 3HJ, UK;*

## ABSTRACT

The Wide-field Spectroscopic Telescope (WST) is a planned 12-meter class dedicated spectroscopic facility for massive spectroscopic surveys. This paper presents the current status of Work Package 4.5, the High Resolution Multi-Object Spectrograph (HR-MOS) module. We describe the international team organization and optical design resulting from extensive trade-off studies, presenting its evolution driven by scientific requirements and technical constraints. Design parameters derived from science cases and astronomical community requirements are detailed. Given the critical importance of mass and volume budgets, we present envelope dimensions and mass estimates for HR-MOS. The spectrograph constructive parameters are defined, including optical fiber specifications, multiplex capability, and modular architecture. Finally, we present the structural analysis addressing mechanical stability and performance requirements for this high-resolution multi-object spectrograph.

**Keywords:** high-resolution spectroscopy, multi-object spectrograph, WST, trade-off study, optical design, astronomical instrumentation, fiber optics, Nasmyth platform

## 1. INTRODUCTION

The Wide-field Spectroscopic Telescope (WST) is a proposed dedicated 12-meter class spectroscopic facility designed to conduct large-scale, multi-object spectroscopic surveys of the sky.[1] Within the WST instrument suite, the High-Resolution Multi-Object Spectrograph (MOS-HR), developed under Work Package 4.5 (WP4.5), is designed to simultaneously observe a large number of targets at resolving powers R = 40,000 over a wavelength range spanning from the near-ultraviolet to the red. This capability addresses a broad spectrum of astrophysical science cases, including characterization of planetary system host stars, and precision cosmology.

The MOS-HR module must satisfy stringent and in part competing requirements: achieving high multiplexing (≥ 2000 targets), high spectral resolution (R ≥ 40,000), broad wavelength coverage in four simultaneous bands, and fitting within the mass and volume budget of the rotating Azimuth floor. This combination drives the system towards a highly modular architecture built around a fiber-slicing concept and a series of identical spectrograph units.

This paper is organized as follows. Section 2 describes the team organization. Section 3 details the top-level scientific requirements. Section 4 presents the optical designs explored. Section 6 covers mechanical design and FEA. Section 7 addresses fiber interface and multiplexing. Section 8 summarizes conclusions and next steps.

## 2. SCIENTIFIC REQUIREMENTS

The top-level requirements for MOS-HR are extracted from the WST Phase A High Level Requirements document.[5] The most relevant requirements driving the optical and mechanical design are:

- **INS-5 – Multiplexing:** ≥ 2000 simultaneous targets per field, achieved by deploying multiple identical spectrograph modules.

[*] andrea.tozzi@inaf.it; phone +390552752282; www.arcetri.inaf.it

• **INS-6 – Resolving power:** R ≥ 40,000, averaged over the covered wavelength range and field position.

• **INS-7 – Wavelength range:** 370–970 nm, in four simultaneous bands each spanning ≈λ/10. Adopted spectral windows: [4009–4431] Å, [4522–4998] Å, [5424.5–5995.5] Å, [6080–6720] Å.

• **INS-9 – Pixel size:** 10×10 µm² or 15×15 µm².

• **INS-10 – Spectral sampling:** ≥ 2.4 pixels per resolution element (minimum value).

An additional cross-talk requirement (private communication) specifies an inter-spectrum spacing of at least twice the fiber core diameter. For a 12-m telescope at f/3.34 with fiber f/# = 3.0, 1 arcsecond on sky projects onto a 194-µm fiber. After a factor-7 slicing, each sub-fiber core is 73 µm in diameter — the two primary dimension drivers of the entire optical design.

## 3. OPTICAL DESIGN

### 3.1 Architecture and fiber-slicing concept

A monolithic single-spectrograph design for 2000 targets at R ≥ 40,000 is not manufacturable with current technology: the required grating sizes would far exceed industrial capabilities. A modular architecture of multiple identical spectrograph units, each serving a fiber sub-set, is therefore mandatory. Figure 1 shows a 3D schematic of a prelimanry and tentative design based on single module as in MOONS, but not feasable in this situation due to the huge dimensions of grating and the overall volume, not compatible with the WST nasmyth platform. For this reason a new design is necessary, as shown in the current paper: the baseline idea (see in Figure 2) is based in a four arms spectrographs, eigth identical modules, but other possible solutions can be found. The baseline design is a good compromise in terms of dimensions, feasibility, manufacturability, cost and expertise of the team.

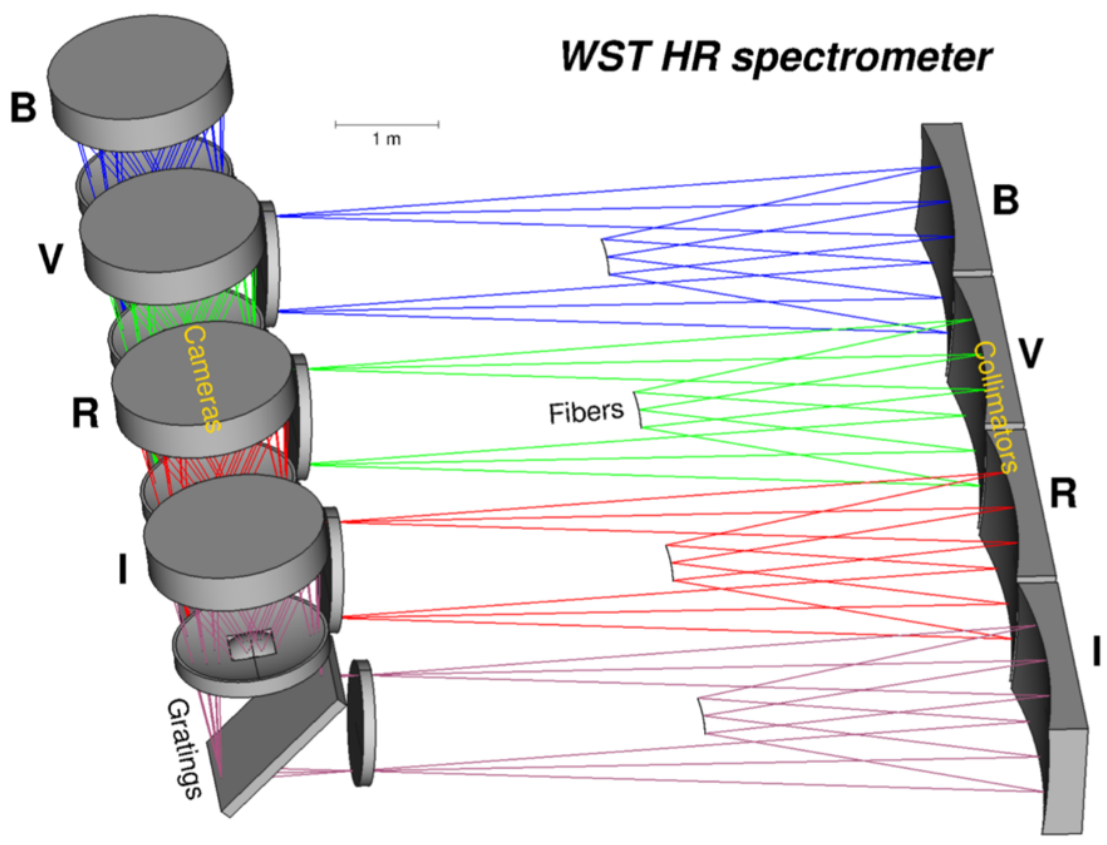

Figure 1. Three-dimensional schematic of a single MOS-HR spectrograph module showing the pseudo-slit (center), spherical-mirror collimator, three dichroics, four Volume Phase Holographic Gratings and the four dioptric cameras for the B (blue), V (green), R (red) and I (pink) spectral bands. Scale bar: 1 m.

In WST the f/# of the telescope id equal to 3.34 for 12 m diameter and so the light from each target is collected as a 194-µm beam and split into sub-fibers through one of two slicing concepts:

**1) Focus Slicer:** 73-µm core fibers placed directly at the telescope focal plane in a hexagonal arrangement. Coupling efficiency ≈63%. This is the baseline approach.

**2) Fiber Slicer (ANDES heritage):** a micro-lens collimates the 194-µm beam and a lenslet array splits it into 7 secondary fibers (modules < 15 cm × < 5 cm diameter). Coupling efficiency ≈58%.

In both cases the spectrograph is agnostic to the slicing method: the optical object for MOS-HR is always the 73-µm fiber core. The pseudo-slit is 266 mm long with a convex curvature radius of 1439 mm; fiber pitch is 100 µm center-to-center (TBC): this "pseudo slit" is the object of the MOS-HR optical design.

### 3.2 Baseline design: on-axis, folded, dioptric camera

The baseline design, proposed by one of the authors B.Delabre, is shows in figure 2 (referred to 8M16D as it is composed by 8 identical modules and by 16 detectors each module) employs an on-axis collimator based on a spherical mirror and an aspheric corrector plate, folded by a flat mirror. The folding mirror is pierced by a rectangular shaped chamfer to accommodate the pseudo-slit. Three dichroics create four spectral channels, each equipped with a VPHG at 45° AoI and a dioptric camera. The Zemax optical layout is shown in Figure 2 and the principal parameters are in Table 1.

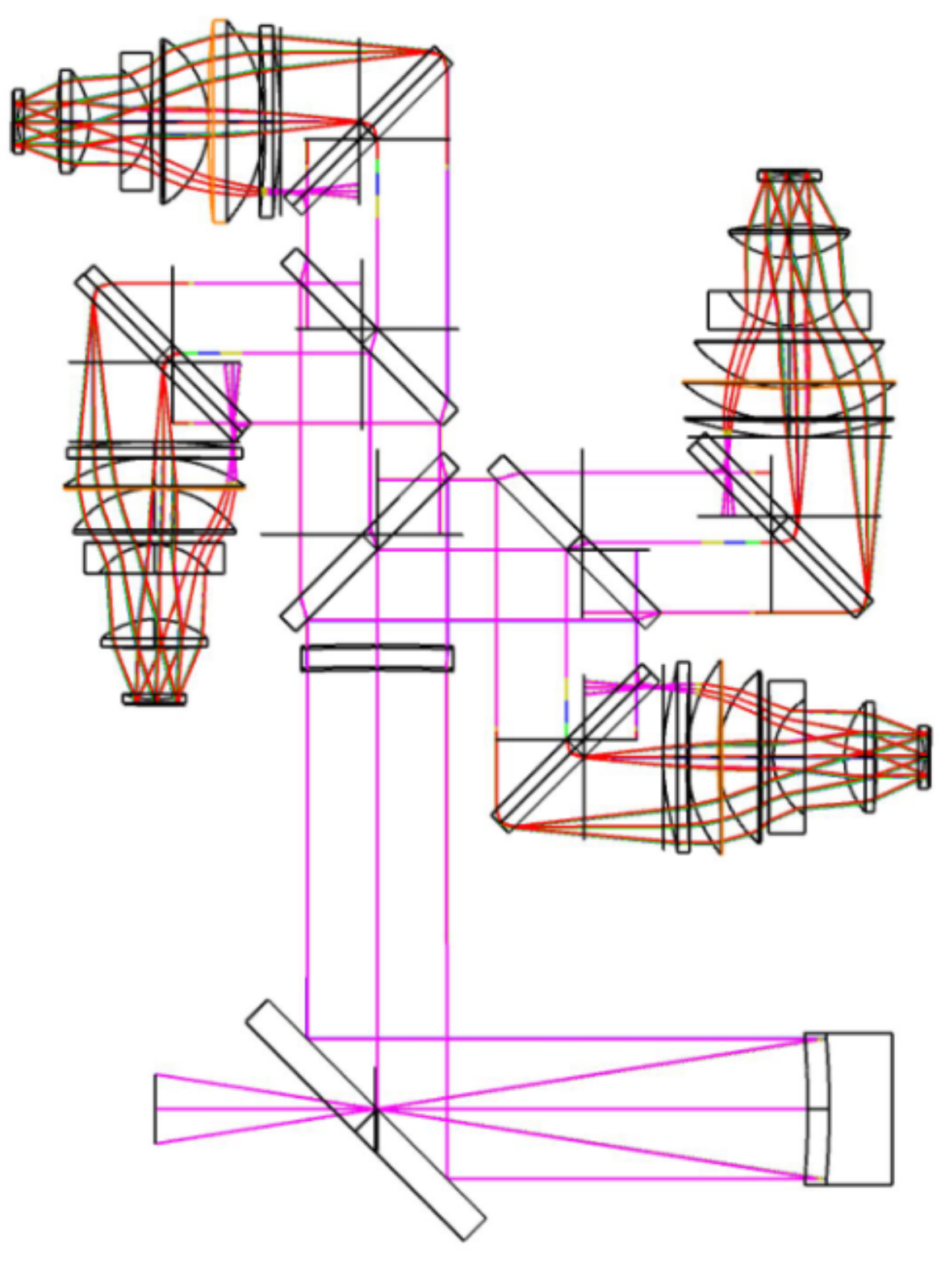

Figure 2. Zemax optical layout of the baseline MOS-HR design (on-axis, folded, dioptric camera). The collimator spherical mirror (bottom) is folded by the flat mirror. The four channels (B blue, V/G yellow-green, R red, I pink) each feature a VPHG and a 6-lens dioptric camera. Scale bar: 2000 mm. This solution is the one called 8M16D: 8 identical moduls, for 16 detectors each module. Total input fibers number around 2500-2800.

In figure 3 is shown a the overall concept idea of the MOS-HR, based on identical modules. The light coming from the scientific objects, let say 250 for each module, is splitted into seven fibers: with a 12 meter diameter telescope and working at f/3.34 we obtain 194 micron for one arcsec in sky. Using F/3.3 fiber optics we obtain a core diameter of 73 micron on the pseudo-slit fibre output. How the X7 splitting is realized, is under discussion: we can have a pupil slicer based on auxiliary optics or we can think about the possibility to put directly the seven 73 micron fibres core, directly on the telescope focal plane: the difference in terms of throughput is quite similar, but the decision involves the positioner, the fibre-link work package and the fibre routing to be compliant with the elevation angle. It is a system level decision that is actually under discussion within the consortium.

In the same figure 3 is shown schematically, the interface between the MOS-HR module and the fiber-link one: it is based on a certain number of connectors disposed in a panel that will be part of the spectrograph itself: in figure there are represented 250x7 science fibers plus one-two fibers between each group of seven. The exact number of fibers between each group of seven, depends by the crosstalk and by the length of the slong slit that actually is something near 266 mm but can reach easily 300 mm: this extra length is taken into account because of the clad and buffer thickness of the fiber manufacturer.

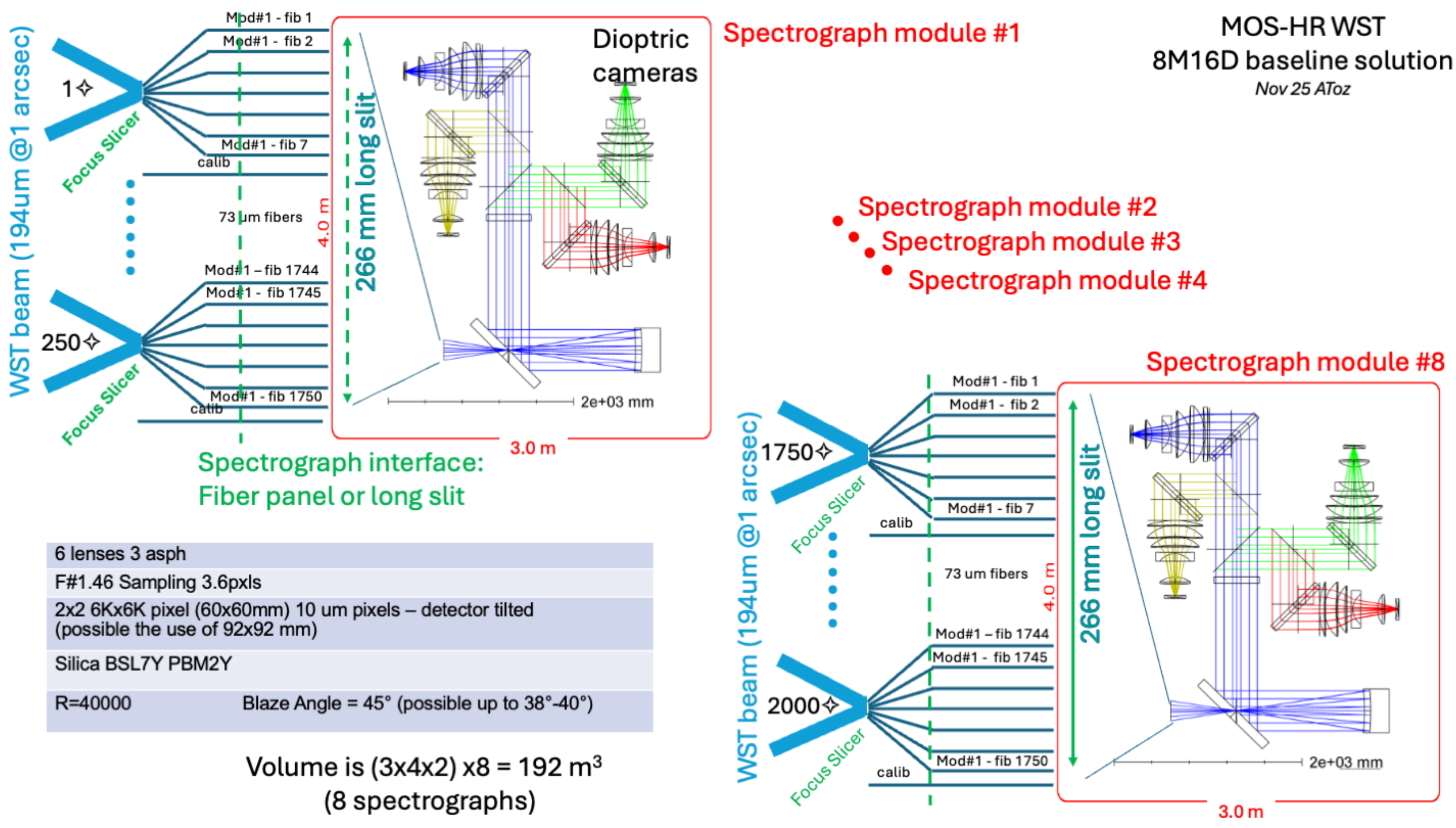

Figure 3. Schema of the MOS-HR modules. Each module is capable to accept 1700 fibers included scientific objects and calibration sourses. The number of modules can be tuned accordingly to the overall mass and volume available in the Azimuth floor. The slicers can be realized as “focus slicer” of “fiber slicer”: in figure is shows schematically the Focus slicer solution.

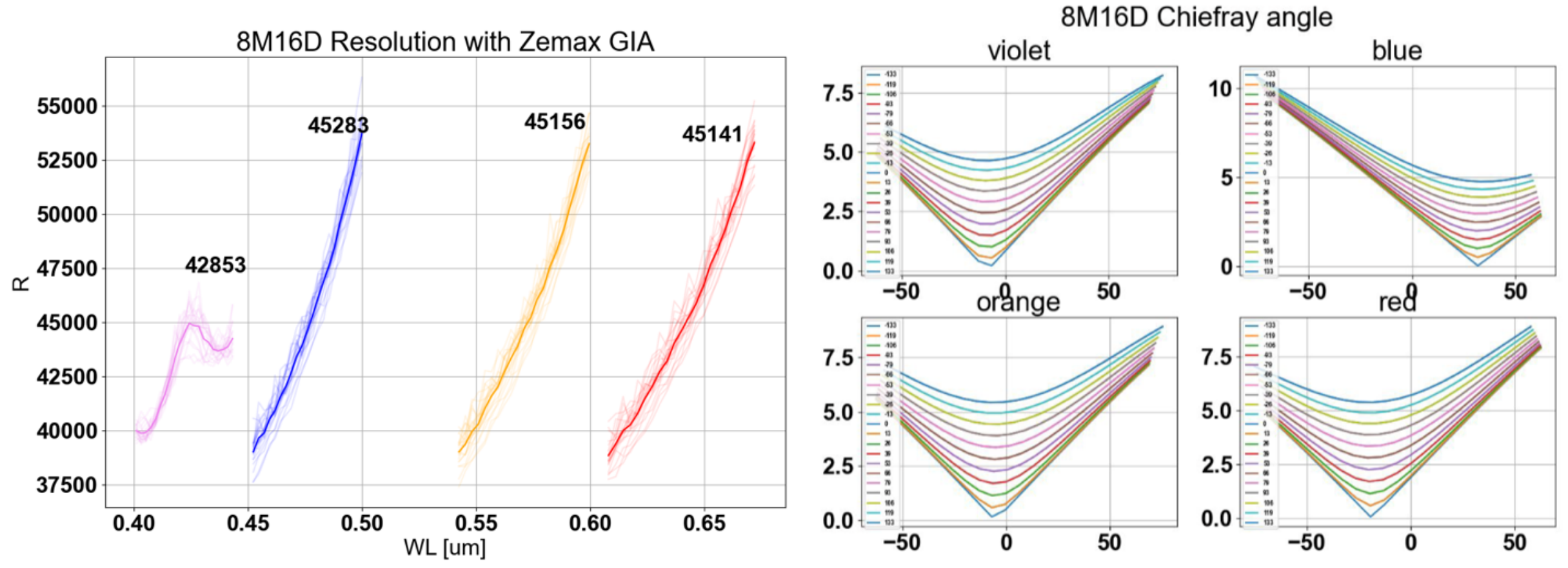

Figure 4. Left: resolution of the MOS-HR modules and on the rigth, the incidence angles on the detecto for the four different arms, calculated using a custom procedure based on the geometric Image Analysis.

The ray angle of incidence at the detector ranges from 0° to 28° for the large design (F/1.46 camera) and from 0° to 38° for the small design (F/1.7 camera). These values are being communicated to the detector group for QE angular response assessment.

Table 1. Baseline design (on-axis, folded, dioptric camera) principal parameters.

| Parameter | Value |
|---|---|
| Pseudo-slit length | 266 mm |
| Pseudo-slit curvature radius | 1439 mm (convex) |
| Collimator spherical mirror | 1250 × 500 mm |
| Collimator folding flat mirror | 1000 mm diameter |
| Collimator aspherical corrector | 710 mm diameter (fused silica) |
| Dichroic size (per channel) | 460 × 660 mm |
| Grating size (per channel) | 460 × 660 mm, AoI = 45° |
| Camera glasses | BSL7Y, fused silica, PBM2Y |
| Camera aspheric surfaces | 3 |
| Camera maximum lens diameter | 690 mm |
| Camera F/# | 1.46 |
| Spectral sampling | 3.0 px / resolution element |
| Detectors per camera | 4 (2×2 mosaic, 6k×6k, 10 μm pixels) |
| Last surface–detector distance | 8 mm (TBC) |
| Envelope per spectrograph unit | 18 $m^3$ |

Resolution is computed as $R = \lambda/\delta\lambda$, with $\delta\lambda$ the FWHM of the geometric image (Zemax Gemetric Image Analysis tool, 1 μm grid) on a 20×20 field-wavelength point grid. The average resolving power is $R > 44000$ for all wavelength ad visible in Figure 4 in the left panel, comfortably above the INS-6 requirement. While the Angle of Incidence (AOI) of the rays on th edetector is shows in the rigth panel of the same figure. The throughput budget (excluding fibers and detector) is given in Table 2.

Table 2. MOS-HR spectrograph throughput (no fibers, no detector). Baseline design, grating AoI = 45° (upper) and 40° (lower).

| Band | Blue (390 nm) | Green (450 nm) | Yellow (620 nm) | Red (700 nm) |
|---|---|---|---|---|
| Mean eff. (AoI=45°) | 0.476 | 0.490 | 0.496 | 0.497 |
| Max eff. (AoI=45°) | 0.508 | 0.515 | 0.511 | 0.525 |
| Mean eff. (AoI=40°) | 0.515 | 0.529 | 0.544 | 0.544 |
| Max eff. (AoI=40°) | 0.541 | 0.553 | 0.562 | 0.568 |

Reducing AoI from 45° to 40° yields a ≈4–5 percentage-point throughput gain across all bands with marginal impact on resolution and volume, making it a promising post-trade-off optimization.

### 3.3 Possible Design variant

**On-axis, folded, catadioptric camera:** A MOONS-style catadioptric camera at F/1.1 replaces the dioptric one, enabling a single 9k×9k, 10-μm detector per channel (spectral sampling = 2.7 px/element, average R = 42,778). However, the focal plane is buried inside the camera body, making thermal stabilization and detector mounting extremely challenging. A 20% vignetting penalty reduces mean throughput to 39–41% per band.

**On-axis, not-folded:** The folding mirror is removed. Volume increases to ∼21 m³/module with essentially unchanged optical performance.

**Off-axis collimator:** Provides more space around the slit assembly but none of the configurations studied meets the optical quality requirement so far.

**Small (16M4D) design:** An alternative design, proposed by one of the authors W.Saunders, with a smaller collimated beam (185 mm vs 440 mm) and 16 modules instead of 8, can be intended as back-up possible silution. Overall azimuth platform footprint is similar even if mass and volume of the single module is approximately half those of the large design (see Figure 5). A possible difficulty into splitting the focal plane fibre of 194 um into 19, that is mandatory for this design to be compliant with the imposed resolving power, is a problem that is under evaluation by fiber link team.

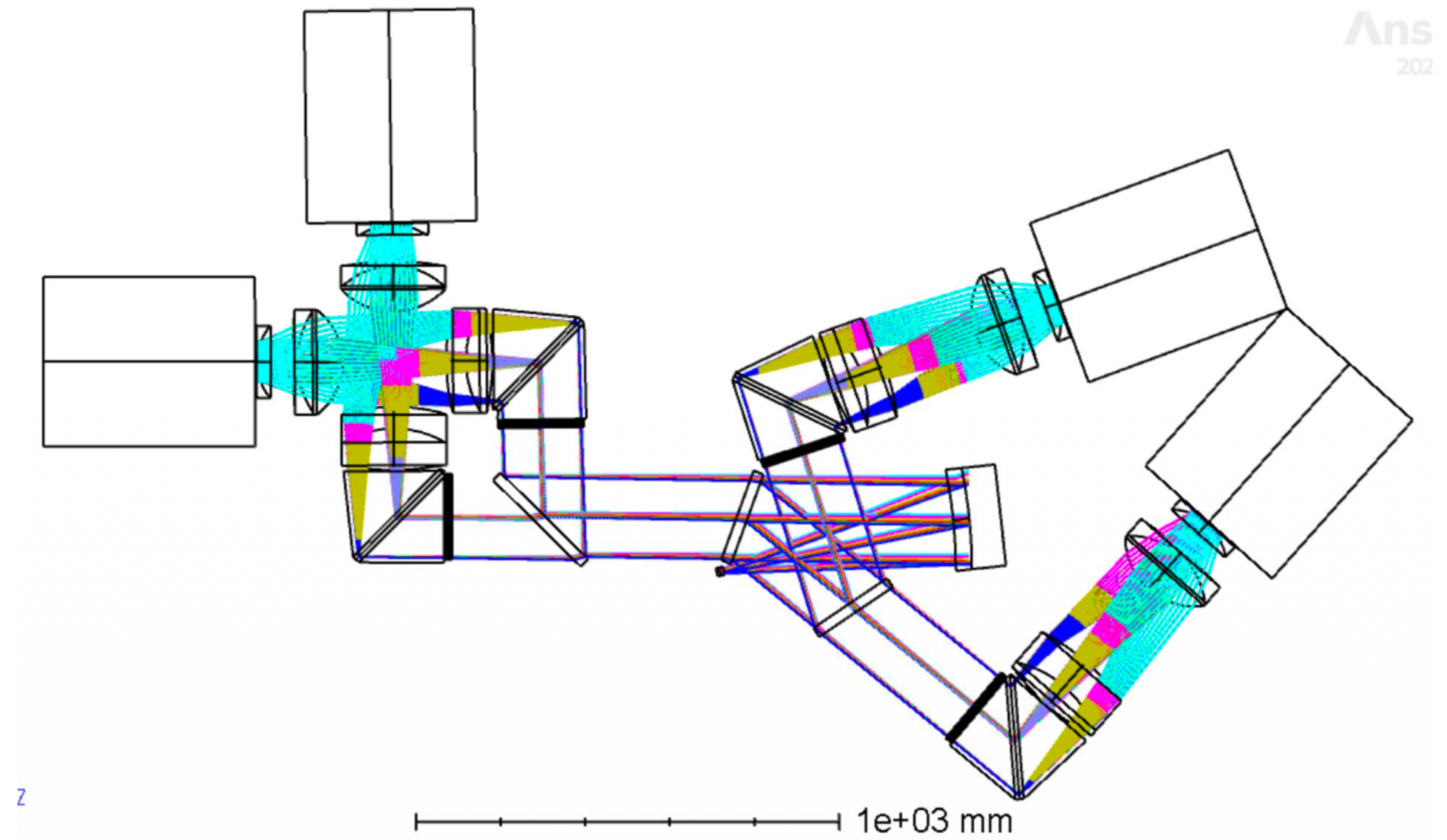

Figure 5. Optical layout of the 16M4D (Thanks to Will Saunders) design showing three spectral channels with the pseudo-slit and collimator. The overall envelope is approximately 3.0 × 4.0 m per module, roughly half the volume of the large baseline design. This solution si the one called 16M4D, that it means 16 identical modules, four detectors each module.

Both designs are being evaluated by the same industrial partners to evaluate the cost, manufacturing complexity of the optics and AIV.

## 4. MECHANICAL DESIGN AND MASS/VOLUME BUDGET

### 4.1 Modular architecture and platform layout

The mechanical design and analysis is deeply described in [8], but here a brief description is furnished.

Six to eight spectrograph modules are required to reach the multiplexing target of 2000 targets (see Section 7). Modules are paired on a shared vertical optical bench forming four double blocks. The vertical configuration provides a compact layout and balanced mass distribution. Figure 6 shows the CAD model of a single unit with an operator for scale.

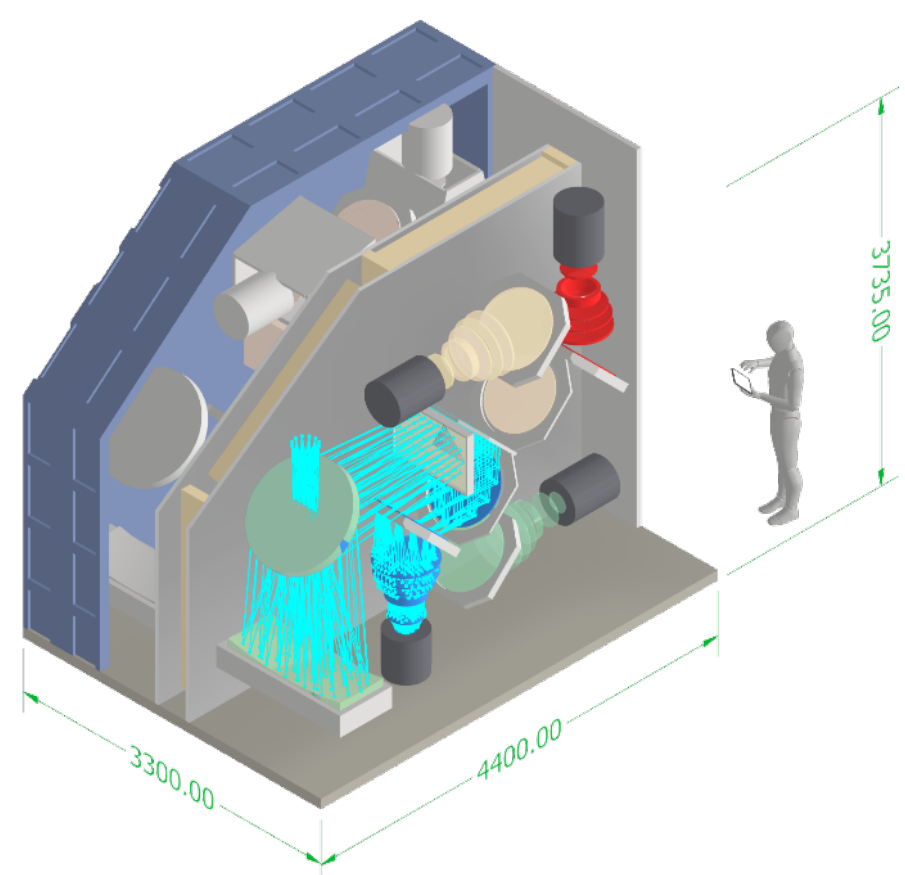

Figure 6. CAD model of a single MOS-HR spectrograph unit in its preliminary mechanical enclosure (3600 × 4400 × 3300 mm, dimensions in mm). An operator figure is included for scale. The colored elements represent the four camera assemblies (dioptric, one per spectral band) mounted on the vertical optical bench.

Each double block has an estimated volume of ~52 $m^3$ and a gross mass of ~21 tonnes (gross coefficient 1.3, non-linearly scaled from MOONS camera heritage). Four blocks total ≈200 $m^3$ and ≈85 tonnes on the rotating Nasmyth platform: in figure 7 is shown a possible solution for the disposition of these eigth modules.

Detector cryostats (~40×40×50 $cm^3$ each) remove approximately 12 W per camera using SunPower Cryotel GT coolers, consistent with DESI and MUSE heritage. Mosaic detector alignment tolerances are estimated at ≈5 µm in focus (Z-shift) and ≈$5\times10^{-3}$ degrees in tilt.

Table 3. MOS-HR mass and volume envelope (8-module, 4-block configuration, baseline design).

| **Parameter** | **Value** |
|---|---|
| Volume per spectrograph unit | 18 $m^3$ (folded) / 21 $m^3$ (not-folded) |
| Volume per double block (2 units) | ≈52 $m^3$ |
| Total volume (4 blocks, 8 units) | ≈200 $m^3$ |
| Gross mass per double block | ≈21 tonnes |
| Total gross mass (4 blocks) | ≈85 tonnes |
| Cryostat volume per camera | 40 × 40 × 50 $cm^3$ |
| Thermal load per camera | ≈12 W (DESI/MUSE heritage) |
| Cryocooler baseline | SunPower Cryotel GT |

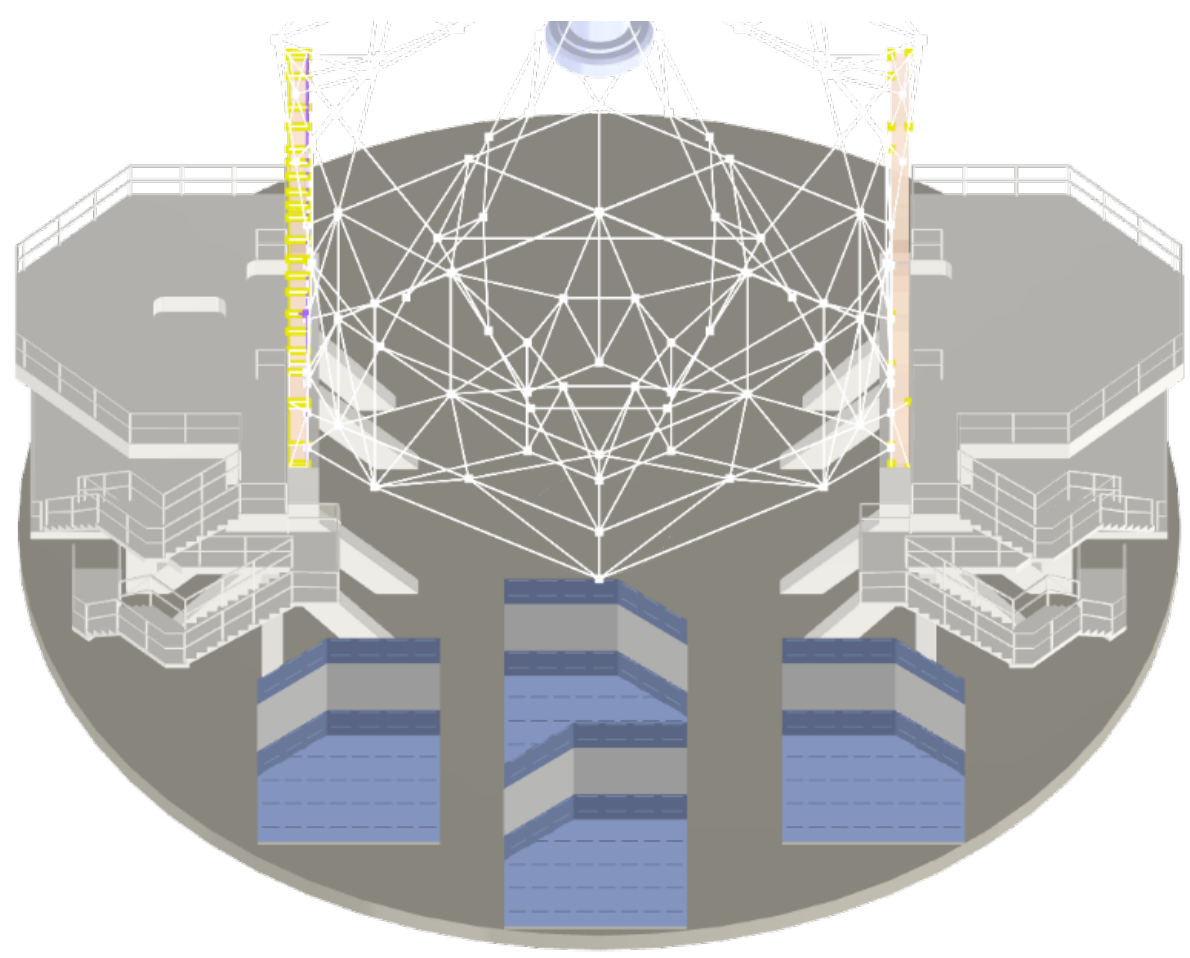

Figure 7. CAD model of a possible layout of 4 double modules of the MOS-HR spectrograph unit on the rotating platform

### 4.2 Structural analysis

A preliminary Finite Element Analysis (FEA) has been performed on the double-block assembly. The structural model uses aluminum as the primary material. The first eigenfrequency of ≈10 Hz is regarded as acceptable for a spectrograph on a rotating platform and is the best value among all configurations. Less compact designs are expected to show lower eigenfrequencies. The FEA will be refined as the design matures, incorporating platform interface loads, thermal gradients, and gravity vector variation with telescope altitude. See [8] for details.

## 5. FIBER INTERFACE AND MULTIPLEXING

The optical interface between the fiber feed system and each MOS-HR spectrograph consists of a fiber panel hosting male-to-male connector terminations. The slit plate has an ultra-low profile (< 2.5 mm). The relationship between the number of required spectrograph modules and the fiber-per-object (slicer) factor is shown in Figure 10.

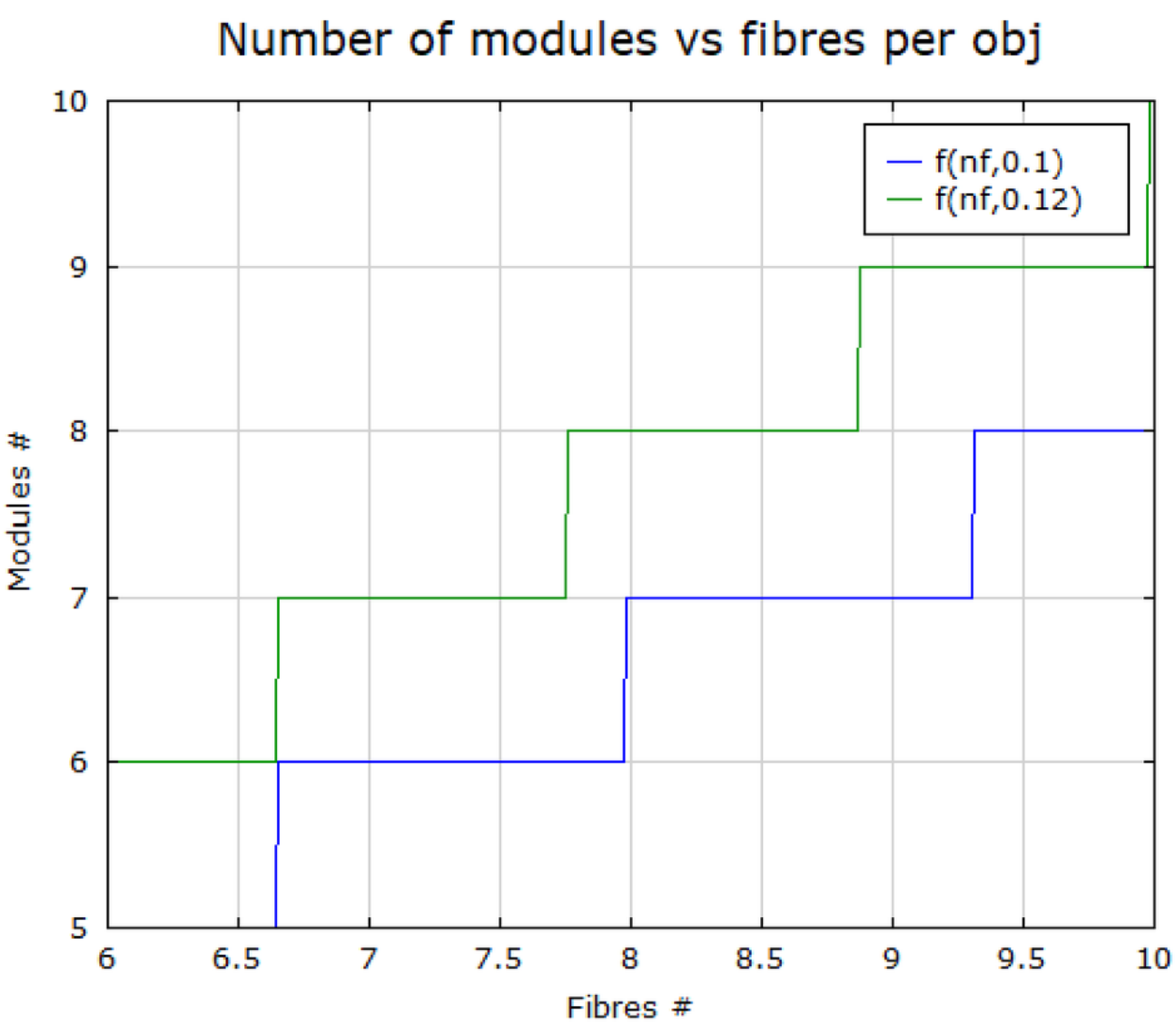

Figure 9. Required number of spectrograph modules to reach ≥ 2000 targets as a function of the number of fibers per object (slicer factor), for fiber pitches of 100 µm (blue) and 120 µm (green). One dark fiber per group is assumed for cross-talk separation. With 7 fibers per target and 100 µm pitch, 6–7 modules are sufficient.

With a 266-mm slit, 100-µm fiber pitch, 7 sub-fibers per target, and one dark fiber between groups, each module accommodates ≈332 targets. Six modules yield ≈1992 targets, meeting the multiplexing requirement. With two dark fibers (spacing-to-core ratio ≈3.1), 7 modules are required for ≈2065 targets. An extension of the slit to 300 mm (larger detector format, 92×92 mm) is being studied to accommodate calibration fibers without reducing science multiplex.

The fiber-slit interface specification — slit length, curvature, fiber count, connector layout — is being defined jointly with the fiber feed WP. Definition of the distance between the slit and the first collimator element (currently no optic is placed in front of the fibers in the large design) and the back-illumination mechanism design are ongoing activities.

## 6. CONCLUSIONS

WP4.5 has made significant progress in defining the MOS-HR design for WST. The trade-off study, structured across six categories with more than 20 weighted parameters, clearly identifies the on-axis, folded, dioptric camera configuration as the preferred baseline. This design achieves an average resolving power R = 42,686, mean throughput 47–50% (AoI=45°, no fiber and detector contributions), first eigenfrequency ≈14.6 Hz (preliminary FEA), and a total system footprint of ≈200 $m^3$ and ≈85 tonnes in an 8-module, 4-block configuration.

A complementary trade-off is in progress to evaluate the large (440 mm collimated beam, 8 modules) versus small (185 mm, 16 modules) designs. The selected baseline will be formally adopted before the next project milestone.

Ongoing activities include: optical optimization of grating AoI and wavelength ranges; cross-talk modeling; mosaic detector alignment tolerance definition; fiber back-illumination mechanism design; cryostat envelope definition; and preparation of the fiber-slit interface specification in coordination with the fiber feed WP.

## ACKNOWLEDGMENTS

The authors acknowledge all members of the WST consortium and the WST Project Office for fruitful discussions and collaborative contributions. The Florence busy-week meeting (October 2025), attended by B. Delabre, is gratefully acknowledged for providing the foundational insights that led to the current baseline design. The authors acknowledgments to the European funds (grant agreement) using the usual sentence and to the "INAF Large Grant 2023 WST: the Wide-dield Spectroscopic Telescope".